\documentclass[preprint]{aastex}

\def\kpc{\mbox{kpc}}
\def\kpch{\mbox{$h^{-1}$kpc}}

\def\Mpch{\mbox{$h^{-1}$Mpc}}

\def\Msunh{\mbox{$h^{-1}$M$_\odot$}}

\def\etal{et al.}
\def\Vmax{\mbox{$V_{\rm max}$}}
\def\rs{\mbox{$r_{\rm s}$}}
\def\rvir{\mbox{$r_{\rm vir}$}}
\def\Mvir{\mbox{$M_{\rm vir}$}}
\def\LCDM{{$\Lambda$CDM}}

\def\mathnew{\mathsurround=0pt}
\def\simov#1#2{\lower .5pt\vbox{\baselineskip0pt
    \lineskip-.5pt\ialign{$\mathnew#1\hfil##\hfil$\crcr#2\crcr\sim\crcr}}}

\def\'#1{\ifx#1i{\accent"13\i}\else{\accent"13#1}\fi}

\def\Cnfw{C_{\rm NFW}}
\def\Cmoore{C_{\rm Moore}}
\shorttitle{Resolving Dark Halos}
\shortauthors{Klypin et al.}

\begin{document}

\title{Numerical Simulations in Cosmology III: Dark Matter Halos}

\author{Anatoly Klypin}
\affil{Astronomy Department, New Mexico State University, Box 30001, Department
4500, Las Cruces, NM 88003-0001}

\begin{abstract}
We give a review of different properties of dark matter halos.  Taken
from different publications, we present results on (1) the mass and
velocity functions, (2) density and velocity profiles, and (3)
concentration of halos.  The results are not sensitive to parameters of
cosmological models, but formally most of them were derived for popular
flat \LCDM~ model.  In the range of radii $r=(0.005-1)\rvir$ the density
profile for a quiet isolated halo is very accurately approximated by a
fit suggested by Moore \etal (1997): $\rho\propto
1/x^{1.5}(1+x^{1.5})$, where $x=r/\rs$ and \rs~ is a characteristic
radius. The fit suggested by Navarro \etal (1995) $\rho\propto
1/x(1+x)^2$, also gives a very satisfactory approximation with relative
errors of about 10\% for radii not smaller than 1\% of the virial
radius. The mass function of $z=0$ halos with mass below $\approx
10^{13}\Msunh$ is approximated by a power law with slope $\alpha
=-1.85$. The slope increases with the  redshift. The velocity function of
halos with $\Vmax< 500$~km/s is also a power law with the slope $\beta=
-3.8-4$. The power-law extends  to halos at least down to 10~km/s. It
is also valid for halos inside larger virialized halos. The
concentration of halos depends on mass (more massive halos are less
concentrated) and environment, with isolated halos being less
concentrated than halos of the same mass inside clusters. Halos have
intrinsic scatter of concentration: at $1\sigma$ level halos with the
same mass have $\Delta(\log{c_{\rm vir}})=0.18$ or, equivalently,
$\Delta\Vmax/\Vmax =0.12$. Velocity anisotropyfor both subhalos and
the dark matter is approximated by $\beta(r) =0.15 + 2x/[x^2+4]$,
where $x$ is radius in units of the virial radius.

\end{abstract}  

%=====================

\section{Introduction}

During the last decade there was an increasingly growing interest in
testing predictions of variants of the cold dark matter (CDM) models
at subgalactic ($\lesssim 100{\rm\ kpc}$) scales.  This interest was
first induced by indications that observed rotation curves in the
central regions of dark matter dominated dwarf galaxies are at odds
with predictions of hierarchical models. Specifically, it was argued
(Moore 1994; Flores \& Primack 1994) that circular velocities,
$v_c(r)\equiv[GM(<r)/r]^{1/2}$, at small galactocentric radii
predicted by the models are too high and increase too rapidly with
increasing radius compared to the observed rotation curves. The
steeper than expected rise of $v_c(r)$ implies that the {\em shape} of
the predicted halo density distribution is incorrect and/or that the
DM halos formed in CDM models are too concentrated (i.e., have too
much of their mass concentrated in the inner regions). 

In addition to the density profiles, there is an alarming mismatch in
predicted abundance of small-mass ($\lesssim 10^8-10^9\Msunh$)
galactic satellites and the observed number of satellites in the Local
Group (Kauffmann, White \& Guiderdoni 1993; Klypin et al. 1999; Moore
et al. 1999).  Although this discrepancy may well be due to feedback
processes such as photoionization that prevent gas collapse and star
formation in the majority of the small-mass satellites (e.g., Bullock,
Kravtsov \& Weinberg 2000), the mass scale at which the problem sets
in is similar to the scale in the spectrum of primordial fluctuations
that may be responsible for the problems with density profiles. In the
age of precision cosmology that forthcoming {\sl MAP} and {\sl Planck}
cosmic microwave background anisotropy satellite missions are expected
to bring, tests of the cosmological models at small scales may prove
to be the final frontier and the ultimate challenge to our
understanding of the cosmology and structure formation in the
Universe. However, this obviously requires detailed predictions and
checks from the theoretical side and higher resolution/quality
observations and thorough understanding of their implications and
associated caveats from the observational side. In this paper we focus
on the theoretical predictions of the density distribution of DM halos
and some problems with comparing these predictions to observations.

A systematic study of halo density profiles for a wide range of halo
masses and cosmologies was done by Navarro, Frenk \& White (1996,
1997; hereafter NFW), who argued that analytical profile of the form
$\rho(r) =\rho_s(r/r_s)^{-1}(1+r/r_s)^{-2}$ provides a good
description of halo profiles in their simulations for all halo masses
and in all cosmologies. Here, $r_s$ is the scale radius which, for
this profile corresponds to the scale at which $d\log\rho(r)/d\log
r\vert_{r=r_s}=-2$. The parameters of the profile are determined by
the halo's virial mass $\Mvir$ and {\em concentration} defined as
$c\equiv \rvir/r_s$. NFW argued that there is a tight correlation
between $c$ and $\Mvir$, which implies that the density distributions
of halos of different masses can in fact be described by a
one-parameter family of analytical profiles. Further studies by
Kravtsov, Klypin \& Khokhlov (1997), Kravtsov et al. (1999), Jing
(1999), Bullock et al. (2000), although confirming the $c(\Mvir)$
correlation, indicated that there is a significant scatter in the
density profiles and concentrations for DM halos of a given mass.

Following the initial studies by Moore (1994) and
Flores \& Primack (1994), Kravtsov et al. (1999) presented a systematic
comparison of the results of numerical simulations with rotation
curves of a sample of seventeen dark matter dominated dwarf and low
surface brightness (LSB) galaxies. Based on these comparisons, we
argued that there does not seem to be a significant discrepancy in the
{\em shape} of the density profiles at the scales probed by the
numerical simulations ($\gtrsim 0.02-0.03\rvir$, where $\rvir$ is
halo's virial radius). However, these conclusions were subject to
several caveats and had to be tested. First, observed galactic
rotation curves had to be re-examined more carefully and with higher
resolution. The fact that all of the observed rotation curves used in
earlier analyses were obtained using relatively low-resolution HI
observations, required checks of the possible beam smearing effects.
Also, a possibility of non-circular random motions in the central
regions that could modify the rotation velocity of the gas (e.g.,
Binney \& Tremain 1987, p. 198) had to be considered. Second, the theoretical
predictions had to be tested for convergence and extended to scales
$\lesssim 0.01\rvir$.

Moore et al. (1998; see also a more recent convergence study by
Ghigna et al. 1999) presented a convergence study and argued that mass
resolution has a significant impact on the central density distribution
of halos. They argued that at least several million particles per halo
are required to model reliably the density profiles at scales
$\lesssim 0.01\rvir$.  Based on these results, Moore et al. (1999)
advocated a density profile of the form $\rho(r)\propto
(r/r_0)^{-1.5}[1+(r/r_0)^{1.5}]^{-1}$, that behaves similarly
($\rho\propto r^{-3}$) to the NFW profile at large radii, but is
steeper at small $r$: $\rho\propto r^{-1.5}$. Most recently,
Jing \& Suto (2000) presented a systematic study of density profiles
for halo masses range $2\times 10^{12}\Msunh-5\times 10^{14}\Msunh$.
The study was uniform in mass and force resolution featuring $\sim
5-10\times 10^5$ particles per halo and force resolution of $\approx
0.004\rvir$. They found that galaxy-mass halos in their simulations
are well fitted by profile\footnote{Note that his profile is somewhat
  different than profile advocated by Moore et al., but behaves
  similarly to the latter at small radii.}  $\rho(r)\propto
(r/r_0)^{-1.5}[1+r/r_0]^{-1.5}$, but that cluster-mass halos are well
described by the NFW profile, with logarithmic slope of the density
profiles at $r=0.01\rvir$ changing from $\approx -1.5$ for $\Mvir\sim
10^{12}\Msunh$ to $\approx -1.1$ for $\Mvir\sim 5\times
10^{14}\Msunh$.  Jing \& Suto interpreted these results as evidence
that profiles of DM halos are not universal.

Rotation curves of a number of dwarf and LSB galaxies have recently
been reconsidered using H$\alpha$ observations (e.g., Swaters, Madore
\& Trewhella 2000; van den Bosch et al. 2000). The results show that
for majority of galaxies H$\alpha$ rotation curves are significantly
different in their central regions than the rotation curves derived
from HI observations. This indicates that the HI rotation curves are
affected by beam smearing (Swaters et al. 2000).  It is also possible
that some of the difference may be due to real differences in
kinematics of the two tracer gas components (ionized and neutral
hydrogen).  Preliminary comparisons of the new H$\alpha$ rotation
curves with model predictions show that NFW density profiles are
consistent with the observed {\em shapes} of rotation curves (van den
Bosch 2000; Navarro \& Swaters 2000).  Moreover, cuspy density
profiles with inner logarithmic slopes as steep as $\sim -1.5$ also
seem to be consistent with the data (van den Bosch 2000).  Nevertheless, CDM
halos appear to be too concentrated (Navarro \& Swaters 2000; McGaugh
et al.  2000; Navarro \& Steinmetz 2000) as compared to galactic
halos, and therefore the problem remains.

New observational and theoretical developments show that comparison of
model predictions to the data is not straightforward. Decisive
comparisons require reaching convergence of theoretical predictions
and understanding the kinematics of the gas in the central regions
observed galaxies. In this paper we present convergence tests designed
to test effects of mass resolution on the density profiles of halos
formed in the currently popular CDM model with cosmological constant
(\LCDM) and simulated using the multiple mass resolution version of
the Adaptive Refinement Tree code (ART). We also discuss some caveats
in drawing conclusions about the density profiles from the fits of
analytical functions to numerical results and their comparisons to the
data. 

\section{Dark Matter Halos: the NFW and the Moore et al. profiles}

Before we fit the analytical profiles to real dark matter halos or
compare them with observed rotational curves, it is instructive to
compare different analytical approximations. Although the NFW and
Moore et al. profiles predict different behavior of $\rho(r)$ in the
central regions of a halo, the scale where this difference becomes
significant depends on the specific values of halo's characteristic
density and radius.  Table~2 presents different parameters and
statistics associated with the two analytical profiles. For the NFW
profile more information can be found in Klypin et al. (1999), Lokas
\& Mamon (2000), and in Widrow (2000). 

Each profile is set by two independent parameters. We choose these to
be the characteristic density $\rho_0$ and radius $r_s$. In this case
all expressions describing different properties of the profiles have
simple form and do not depend on concentration.  The concentration or
the virial mass appear only in the normalization of the expressions.
The choice of the virial radius (e.g., Lokas \& Mamon 2000) as a scale
unit results in more complicated expressions with explicit dependence
on the concentration. In this case, one also has to be careful about
definition of the virial radius, as there are several different
definitions in the literature. For example, it is often defined as the
radius, $r_{200}$, within which the average density is 200 times the
{\em critical density}. In this paper the virial radius is defined as
radius within which the average density is equal to the density
predicted by the top-hat model: it is $\delta_{\rm TH}$ times the {\em
  average matter density} in the Universe.  For the $\Omega_0=1$ case
the two existing definitions are equivalent.  In the case of
$\Omega_0=0.3$ models, however, the virial radius is about 30\% larger
than $r_{200}$.

\begin{figure}[tb!]
\epsscale{0.7}
\plotone{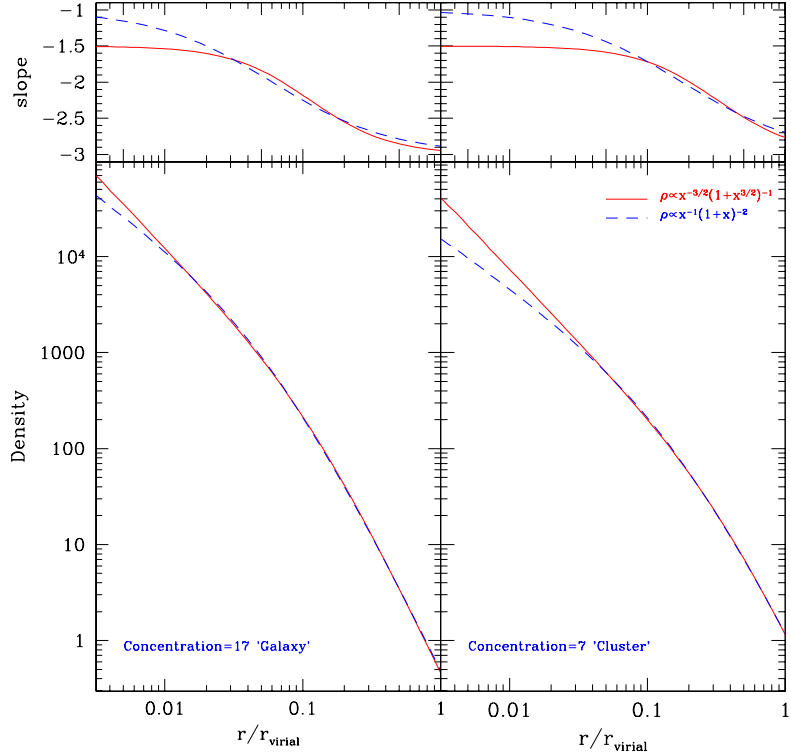}
\caption{\small Comparison of the Moore et al. and the NFW profiles. 
  Each profile is normalized to have the same virial mass and the same
  radius of the maximum circular velocity.  {\it Left panels:} High
  concentration halo with concentration typical for small galaxies
  $C_{\rm NFW}=17$ {\it Right panels:} Low concentration halo with
  concentration typical for clusters of galaxies. The deviations are
  very small ($<3\%$) for radii $r>1/2r_s$. Top panels show the local
  logarithmic slope of the profiles. Note that for the high
  concentration halo the slope of the profile is significantly larger
  than the asymptotic value -1 even at very small radii $r \approx
  0.01/r_{\rm vir}$.  }\label{fig:CompareBenNFW}
\end{figure}

There is no unique way of defining a consistent concentration for the
different analytical profiles. Again, it is natural to use the
characteristic radius $r_s$ to define the concentration: $c\equiv
r_{\rm vir}/r_s$.  This simplifies expressions. At the same time, if
we fit the same dark matter halo with the two profiles, we will get
different concentrations because the values of corresponding $r_s$
will be different. Alternatively, if we choose to match the outer
parts of the profiles (say, $r> r_s$) as closely as possible, we may
choose to change the ratio of the characteristic radii $r_{s, \rm
  NFW}/r_{s, \rm Moore}$ in such a way that both profiles reach the
maximum circular velocity $v_{\rm circ}$ at the same physical radius
$r_{\rm max}$. In this case, the formal concentration of the Moore et
al. profile is 1.72 times smaller than that of the NFW profile.
Indeed, with this normalization profiles look very similar in the
outer parts as one finds in Figure \ref{fig:CompareBenNFW}.  Table~2
also gives two other ``concentrations''. The concentration $C_{1/5}$
is defined as the ratio of virial radius to the radius, which
encompasses 1/5 of the virial mass (Avila-Reese et al. 1999). For
halos with $C_{\rm NFW}\approx 5.5$ this 1/5 mass concentration is
equal to $C_{\rm NFW}$. One can also define the concentration as the
ratio of the virial radius to the radius at which the logarithmic
slope of the density profile is equal -2. This scale corresponds to
$r_s$ for the NFW profile and $\approx 0.35 r_s$ for the Moore et al.
profile.

\begin{deluxetable}{l|l|l} 
\tablecolumns{3} 
\tablewidth{0pc} 
\tablecaption{Comparison of  NFW and Moore et al. profiles} 
\tablehead{ 
\colhead{Parameter} & \colhead{NFW}   & \colhead{Moore et al.}  
} 
\startdata 
Density &  $\rho = \frac{\displaystyle \rho_0}{\displaystyle x(1+x)^2}$ & 
 $\rho = \frac{\displaystyle \rho_0}{\displaystyle x^{1.5}(1+x)^{1.5}}$ \\ 
\quad $x=r/r_s$ & $\quad \rho \propto x^{-3}$ for $x\gg 1$ & $\quad \rho \propto x^{-3}$ for $x\gg 1$ \\
                        & $\quad \rho \propto x^{-1}$ for $x\ll 1$ & $\quad \rho \propto x^{-1.5}$ for $x\ll 1$ \\
                 & $\quad \rho/\rho_0 =1/4\phm{.00}$ at $x=1$ &  $\quad \rho/\rho_0 =1/2\phm{.00}$ at $x=1$ \\
                 & $\quad \rho/\rho_0 =1/21.3$ at $x=2.15$ &  $\quad \rho/\rho_0 =1/3.35$ at $x=1.25$ \\ 
\tableline
Mass     & & \\
\quad $M=4\pi\rho_0r_s^3f(x)$ & 
 $f(x)= \ln(1+x) -\frac{\displaystyle x}{\displaystyle 1+x}$ & $f(x)= \frac{2}{3}\ln(1+x^{3/2})$ \\
\quad   $\phm{M}=M_{\rm vir}f(x)/f(C)$ & & \\
\quad  $M_{\rm vir}=\frac{4\pi}{3}\rho_{\rm cr}\Omega_0\delta_{\rm th}r_{\rm vir}^3$ & \\
\tableline
Concentration &$\Cnfw = 1.72\Cmoore$   & $\Cmoore = \Cnfw/1.72$   \\
\quad $C=r_{\rm vir}/r_s$ & \quad (for the same $M_{\rm vir}$ and $r_{\rm max}$) &  \\
                                & $C_{1/5} \approx \frac{\displaystyle \Cnfw}{\displaystyle 0.86f(\Cnfw)+0.1363}$ &
                                              $C_{1/5} = \frac{\displaystyle \Cmoore}{[\displaystyle (1+\Cmoore^{3/2})^{1/5}-1]^{2/3}} $ \\
                                        &\phm{..}(error $<% 3\% for $\Cnfw=5-30$)  & \phm{C0.0}$
                                                     \approx \frac{\displaystyle \Cmoore}{\displaystyle [\Cmoore^{3/10}-1]^{2/3}}$ \\
      & $C_{\gamma=-2}=\Cnfw$ & $C_{\gamma=-2}=2^{3/2}\Cmoore$ \\
      &  & \phm{$C_{\gamma=-2}=$}$\approx 2.83\Cmoore$ \\
\tableline
Circular Velocity &  & \\
\quad $v_{\rm circ}^2 =\displaystyle \frac{GM_{\rm vir}}{r_{\rm vir}}\frac{C}{x}\frac{f(x)}{f(C)}$ &
             $x_{\rm max} \approx 2.15$ &  $x_{\rm max} \approx 1.25$ \\
\quad  $ \phm{v_{\rm circ}^2}=\displaystyle v_{\rm max}^2\frac{x_{\rm max}}{x}\frac{f(x)}{f(x_{\rm max})} $ &
         $v_{\rm max}^2 \approx 0.216v_{\rm vir}^2\displaystyle \frac{C}{f(C)}$&
          $v_{\rm max}^2 \approx 0.466v_{\rm vir}^2\displaystyle \frac{C}{f(C)}$ \\
\quad $v_{\rm vir}^2\phd=\displaystyle \frac{ GM_{\rm vir}}{r_{\rm vir}}$         &  & \\
\enddata 
\end{deluxetable} 

Figure \ref{fig:CompareBenNFW} presents comparison of the analytic
profiles normalized to have the same virial mass and the same radius
$r_{\rm max}$.  We show results for halos of low and high values of
concentration representative of cluster- and low-mass galaxy halos,
respectively. The bottom panels show the profiles, while the top
panels show corresponding logarithmic slope as a function of radius.
The figure shows that the two profiles are very similar throughout the
main body of the halos. Only in the very central region the
differences become significant. The difference is more apparent in the
logarithmic slope than in the actual density profiles. Moreover, for
galaxy-mass halos the difference sets in at a rather small radius
$\lesssim 0.01\rvir$, which would correspond to scales $< 1{\rm\ kpc}$
for the typical dark matter dominated dwarf and LSB galaxies. In most 
analyses involving galaxy-size halos, the differences between NFW and Moore
et al. profiles are irrelevant, and NFW profile should provide an accurate 
description of the density distribution.
 
Note also that for galaxy-size (e.g., high-concentration) halos the
logarithmic slope of the NFW profile does not reach its asymptotic
inner value of $-1$ at scales as small as $0.01\rvir$.
 For $\sim 10^{12}\Msunh$
halos the logarithmic slope of the NFW profile is $\approx -1.4-1.5$,
while for cluster-size halos this slope is $\approx -1.2$. This
dependence of slope at a given fraction of the virial radius on the
virial mass of the halo is very similar to the results plotted in
Figure~3 of Jing \& Suto (2000). These authors interpreted it as
evidence that halo profiles are not universal. It is obvious, however,
that their results are consistent with NFW profiles and the dependence
of the slope on mass can be simply a manifestation of the well-studied
$c_{\rm vir}(M)$ relation.

To summarize, we find that the differences between the NFW and the
Moore et al. profiles are very small ($\Delta\rho/\rho < 10\%$) for
radii above 1\% of the virial radius. The differences are larger for
halos with smaller concentrations. In case of the NFW profile, the
asymptotic value of the central slope $\gamma  =-1$ is not achieved
even at radii as small as 1\%-2\% of the virial radius.  

\section{Properties of Dark Matter Halos}
Some properties of halos depend on large-scale environment in which the
halos are found. We will call a halo {\it distinct} if it is not inside a
virial radius of another (larger) halo. Halo is called {\it subhalo}, if it is
inside of another halo. The number of subhalos depends on the mass
resolution -- the deeper we go, the more subhalos we will find. Most of
the results below are based on a simulation, which was complete to
masses down to $10^{11}\Msunh$ or, equivalently, to the maximum circular
velocity of 100~km/s.

\begin{figure}[tb!]
\epsscale{0.7}
\plotone{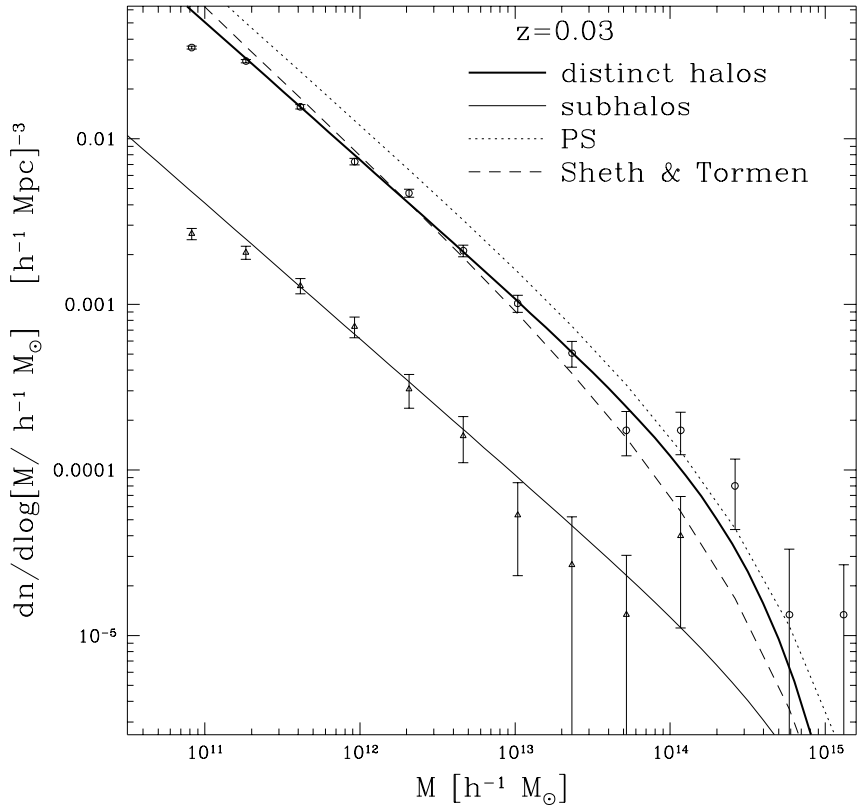}
\caption{\small Mass function for distinct halos (top) and for subhalos
bottom). Raw counts are marked by symbols with error bars. The curves
are Schechter-function fits. Press-Schechter (dotted) and Sheth-Tormen
(dashed) predictions for distinct halos are also shown. On scales
below $10^{14}\Msunh$ the mass function is close to a power law with the slope
$\alpha \approx -1.8$. There is no visible difference in the slope for
subhalos and for the distinct halos. (After
\citet{Sigad})
 }\label{fig:MassFunction}
\end{figure}

{\bf Mass and Velocity Distribution functions}. 
Extensive analysis of halo mass and velocity function was done by
\citet{Sigad} for halos in the \LCDM~ model. Additional results can also
be found in \citet{ghigna99, Moore99b, KKVP99, Gottlober}.
Figure~\ref{fig:MassFunction} compares the mass function of subhalos
and distinct halos. The Press-Schechter approximation overestimates the
the mass function by a factor of 2 for $M< 5\times 10^{12}\Msunh$ and
it somewhat underestimates it at larger masses. More advanced
approximation given by Sheth \& Tormen is more accurate. On scales
below $10^{14}\Msunh$ the mass function is close to a power law with the slope
$\alpha \approx -1.8$. There is no visible difference in the slope for
subhalos and for the distinct halos. 

\begin{figure}[tb!]
\epsscale{0.8}
\plotone{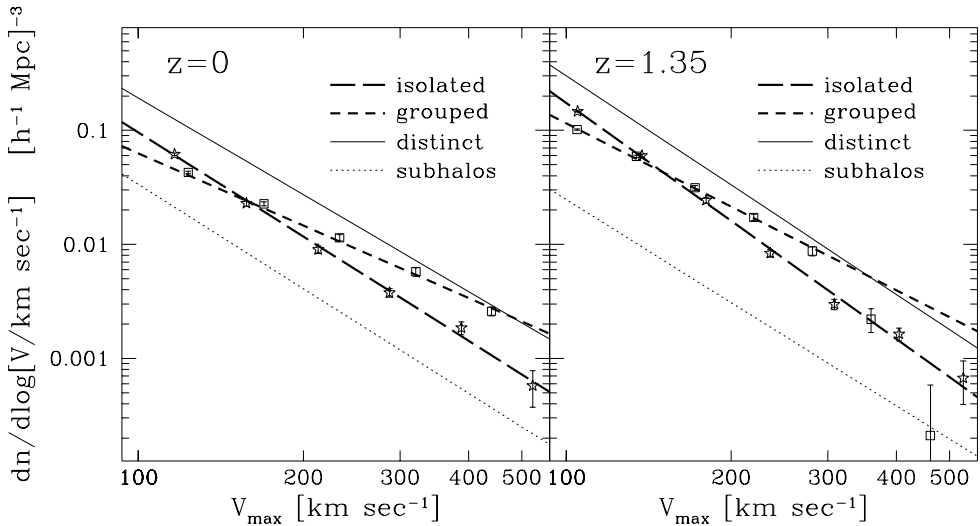}
\caption{\small Velocity functions for isolated halos (squares) and for halos
in groups and clusters. Halos with mass less than $10^{13}\Msunh$ are
used for the plots. (After \citet{Sigad}) }\label{fig:VelocityFunction}
\end{figure}

For each halo one can measure the maximum circular velocity $V_{\rm
max}$. In many cases (especially for subhalos) $V_{\rm max}$ is a
better measure of how large is the halo. It is also closer related with
observed properties of galaxies hosted by
halos. Figure~\ref{fig:VelocityFunction} presents the velocity
distribution function of different types of halos. In addition to
distinct halos and subhalos, we show also isolated halos and halos in
groups and clusters. Here isolated halos are defined as halos with mass
less than $10^{13}\Msunh$, which are not inside a larger halo and which
do not have subhalos more massive than $10^{11}\Msunh$. The velocity
function is approximated by a power law $dn=\Phi_*V_{\rm
max}^{\beta}dV_{\rm max}$ with the slope $\beta\approx -3.8$ for
distinct halos. The slope depends on environment: $\beta\approx -3.1$
for halos in groups and $\beta\approx -4$ for isolated halos.
\citet{KKVP99} and \citet{ghigna99}
found that the slope $\beta\approx -3.8-4$ of the velocity function
extends to much smaller halos with velocities down to 10~km/s. 

{\bf Correlation between characteristic density and radius.}
The halo density profiles are approximated by the Navarro-Frenk-White
profile:

\begin{equation}
\rho =\frac{\rho_0}{(r/r_0)[1+r/r_0]^2}\label{eq:NFW}
\end{equation}

\begin{figure}[tb!]
\epsscale{0.8}
\plotone{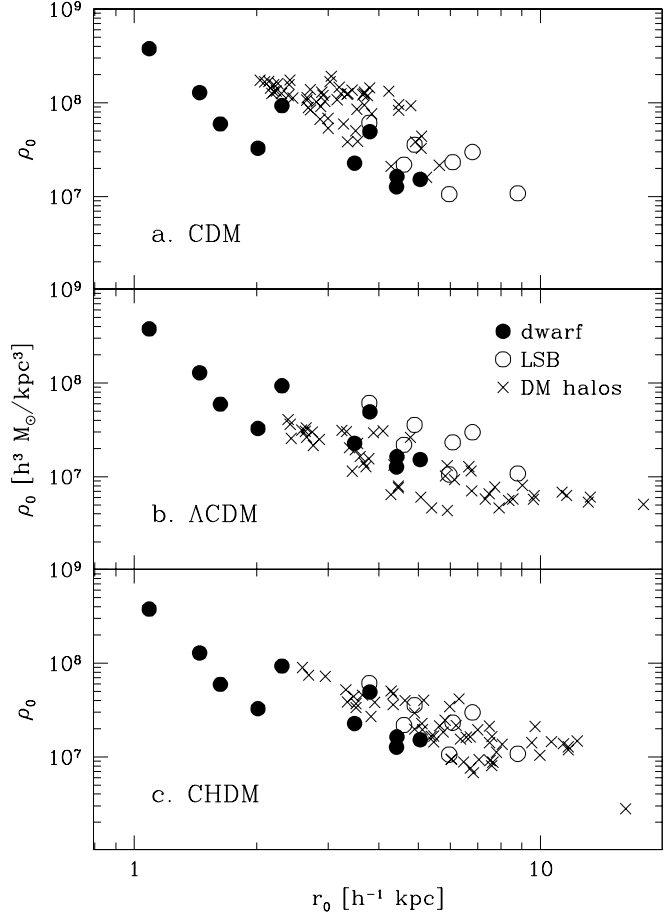}
\caption{\small Correlation of the characteristic density $\rho_0$ and
radius $r_0$ for the dwarf and LSB galaxies (solid and open circles)
and for the dark matter halos (crosses) in different cosmological
models. Halos are consistent with observational data: smaller halos are
denser. (After \citet{KKBP99})}\label{fig:RhoRs}
\end{figure}

\citet{KKBP99} find the correlation between the two parameters of
halos $\rho_0$ and $r_s$. Fifure~\ref{fig:RhoRs} compares results for
the dark matter halos with those for the the dark matter dominated Low
Surface Brightness (LSB) galaxies and dwarf galaxies.  Halos are
consistent with observational data: smaller halos are denser.

{\bf Correlations between mass, concentration, and redshift.}
\citet{nfw97} argued that the halo profiles have a universal shape in the
sense that profile is uniquely defined by virial mass of the halo. 
\cite{Bullocketal99} analyzed concentrations of thousands of halos at
different redshifts. To some degree they confirm conclusions of
\citet{nfw97}: halo concentration correlates with its mass. But some
significant deviations were also found. There is no one-to-one relation
between concentration and mass. It appears the the universal profile
should only be treated as a trend: halo concentration does increase as
the halo mass decreases, but there are large deviations for individual
halos from that ``universal'' shape. Halos have intrinsic scatter of
concentration: at $1\sigma$ level halos with the same mass have
$\Delta(\log{c_{\rm vir}})=0.18$ or, equivalently, $\Delta\Vmax/\Vmax
=0.12$.

\begin{figure}[tb!]
\epsscale{1.05}
\plottwo{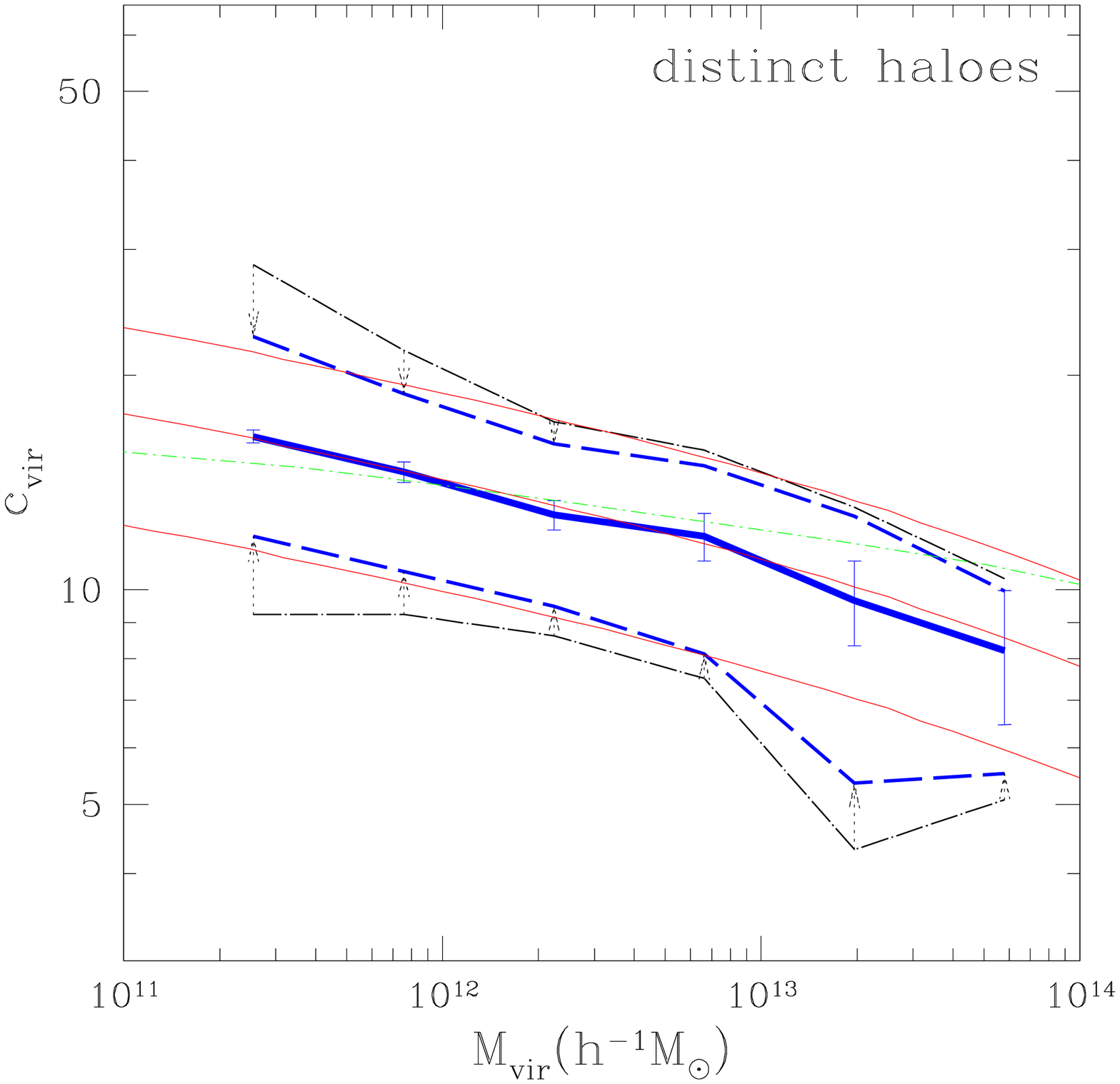}{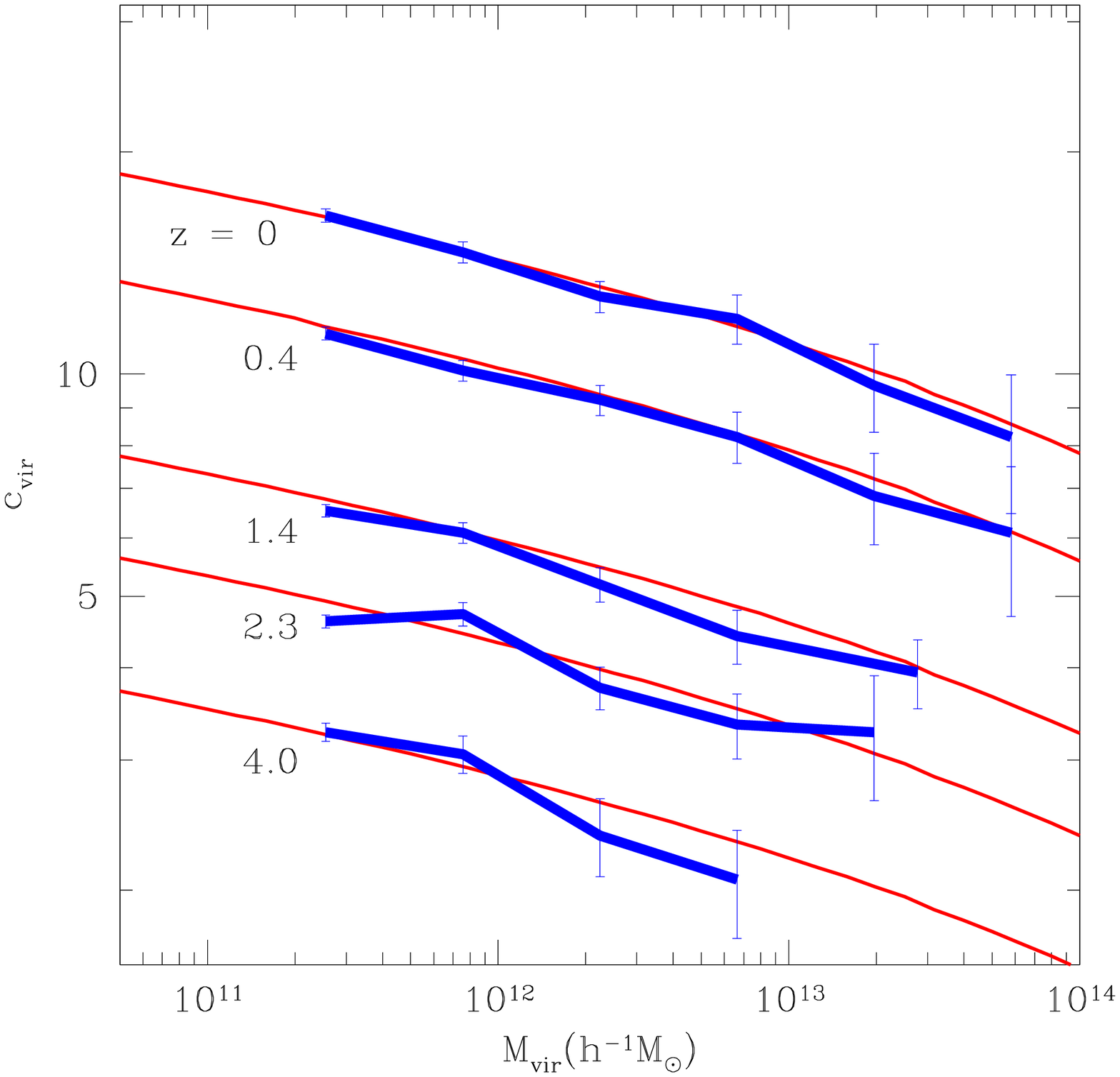}
\caption{\small {\it Left Panel} Dependence of concentration with mass
for distinct halos. The thick solid curve is the median value. The errors
are errors of the mean due to sampling. The outer dot-dashed curves
encompass 68\% of halos in the simulations. The dashed curves and
arrows indicate values corrected for the noise in halo profiles.Thin
curves are different analytical models.  {\it Right Panel} Median halo
concentration as a function of mass for different redshifts. The thin
lines show predictions of an analytical model.
(After \citet{Bullocketal99}). }\label{fig:Cvmass}
\end{figure}

{\bf Velocity anisotropy.} Inside a large halo subhalos or dark matter
particles do not move on either circular or radial orbits. Velocity
ellipsoid can be measured at each position inside halo. It can be
characterized by anisotropy parameter defined as $\beta(r) = 1-
V^2_\perp/2V^2_r$.
Here $V^2_\perp$ is the velocity dispersion perpendicular to the radial
direction  and $V^2_r$ is the radial velocity dispersion. For pure
radial motions $\beta=1$. For isotropic velocities $\beta =0$. Function
$\beta(r)$ was estimated for halos in different cosmological models (
see \citet{Colin99} for references). By studying 12 rich clusters with
many subhalos inside each of them, \citet{Colin99} found that both the
subhalos and dark matter particles can be described by the same
anisotropy parameter

\begin{equation}
\beta(r) =0.15 +\frac{2x}{x^2+4}, \quad x =r/r_{\rm vir}
\end{equation}

\section{Halo profiles: Convergence study}
The following results are based on \citet{KKBP2000}
\subsection{Numerical simulations}
\label{sec:simulations}

\begin{deluxetable}{ccccccccccc} 
\tablecolumns{11} 
\tablewidth{0pc} 
\tablecaption{Parameters of Halos} 
\tablehead{
\colhead{} & \colhead{z}   & \colhead{$M_{\rm vir}$}    & \colhead{$R_{\rm vir}$} & 
\colhead{$V_{\rm max}$}    & \colhead{$N_{\rm part}$}   &
\colhead{$m_{\rm part}$}    & \colhead{Form.res.}    & \colhead{$C_{\rm NFW}$}    & 
\colhead{RelEr}  & \colhead{RelEr} \\
\colhead{} & \colhead{}   & \colhead{$M_{\odot}/h$}    & \colhead{kpc/h} & 
\colhead{km/s}    & \colhead{}   &
\colhead{$M_{\odot}/h$}    & \colhead{kpc/h}    & \colhead{}    & 
\colhead{NFW} & \colhead{Moore} \\
\colhead{(1)} &\colhead{(2)} &\colhead{(3)} &\colhead{(4)} &\colhead{(5)} &\colhead{(6)}
&\colhead{(7)} &\colhead{(8)} &\colhead{(9)} &\colhead{(10)} &\colhead{(11)} 
}
\startdata 
A$_1$ & 0 & $1.97\cdot 10^{12}$ & 257 &247.0 &$1.2\cdot 10^5$   & $1.6\cdot 10^7$ & 0.23 & 17.4& 0.17& 0.20 \\ 
A$_2$ & 0 & $2.05\cdot 10^{12}$ & 261 &248.5 &$1.5\cdot 10^4$   &  $1.3\cdot 10^8$ & 0.91 & 16.0& 0.13& 0.16 \\ 
A$_3$ & 0 & $1.98\cdot 10^{12}$ & 256 &250.5 &$1.9\cdot 10^3$  &$1.1\cdot 10^9$  & 3.66 & 16.6&0.16 & 0.10 \\ 
B         & 1 & $8.5\cdot 10^{11}$ & 241 &195.4 & $7.1\cdot 10^5$ & $1.2\cdot 10^6$ & 0.19 & 12.3& 0.23& 0.16 \\ 
C         & 1 & $6.8\cdot 10^{11}$ & 208 &165.7 & $5.0\cdot 10^5$ & $1.2\cdot 10^6$ & 0.19 & 11.9& 0.37& 0.20 \\ 
D         & 1 & $9.6\cdot 10^{11}$ & 245 &202.4 & $7.9\cdot 10^5$ & $1.2\cdot 10^6$ & 0.19 & 9.5&0.25 & 0.60 \\ 
\enddata 
\end{deluxetable} 
Using the ART code \citep{KKK97, PhD}, we simulate a flat low-density
cosmological model ($\Lambda$CDM) with $\Omega_0 = 1 - \Omega_\Lambda =
0.3$, the Hubble parameter (in units of $100{\rm\ km s^{-1} Mpc^{-1}}$)
$h=0.7$, and the spectrum normalization $\sigma_8=0.9$.  We have run
two sets of simulations with $30\Mpch$ and $25\Mpch$ computational
box. The first simulations were run to the present moment $z=0$.  The
second set of simulations had higher mass resolution and therefore
produced more halos but were run only to $z=1$.

In all of our simulations step in the expansion parameter was chosen
to be $\Delta a_0=2\times 10^{-3}$ on the zero level of resolution.
This gives about 500 steps for an entire run to $z=0$. A test run was
done with twice smaller time-step for a halos of comparable mass (but
with smaller number of particles) as studied in this paper. We did not
find any visible deviations in the halo profile. In the first set of
simulations, the highest level of refinement was ten, which
corresponds to $500\times 2^{10}\approx 500,000$ time steps at the
tenth level.  For the second set of simulation, nine levels of
refinement were reached which corresponds to $128,000$ steps at the
ninth level.

In the following sections we present results for four halos. The first
halo ($A$) was the only halo selected for resimulation in the first
set of simulations. In this case the selected halo was relatively
quiescent at $z=0$ and had no massive neighbors. The halo was located
in a long filament bordering a large void.  It was about 10~Mpc away
from nearest cluster-size halo. After the high-resolution simulation
was completed we found that the nearest galaxy-size halo was about
5~Mpc away.  The halo had a fairly typical merging history with $M(t)$
track slightly lower than the average mass growth predicted using
extended Press-Schechter model.  The last major merger event occured
at $z\approx 2.5$; at lower redshifts the mass growth (the mass in
this time interval has grown by a factor of three) was due to slow and
steady mass accretion.

The second set of simulations was done in a different way. In the low
resolution run we selected three halos in a well pronounced filament.
Two of the halos are neighbors located at about 0.5~Mpc from each
other. The third halo was 2~Mpc away from this pair. Thus, the halos
were not selected to be too isolated as was the case in the first set
of runs. Moreover, the simulation was analyzed at an earlier moment
($z=1$) where halos are more likely to be unrelaxed. Therefore, we
consider the halo $A$ from the first set as an example of a rather
isolated well relaxed halo. In many respects, this halo is similar to
halos simulated by other research groups that used multiple mass
resolution techniques. The three halos from the second set of
simulations can be viewed as representative of more typical halos, not 
necessarily well relaxed and located in more crowded environments.

Parameters of the simulated dark matter halos are listed in Table~2.
Columns in the table present (1) Halo ``name'' (halos A$_1$, A$_2$,
A$_3$ are the halo A re-simulated times with different resolutions);
(2) redshift at which halo was analyzed; (3-5) virial mass, comoving
virial radius, and maximum circular velocity. At $z=0$ ($z=1$) the
virial radius was estimated as the radius within which the average
overdensity of matter is 340 (180) times larger than the mean
cosmological density of matter at that redshift; (6) the number of
particles within the virial radius; (7) the smallest particle mass in
the simulation; (8) formal force resolution achieved in the
simulation. As we will show below, convergent results are expected at
scales larger than four times the formal resolution; (9) halo
concentration as estimated from NFW profile fits to halo density
profiles; (10) maximum relative error of the NFW fit: $\rho_{\rm
  NFW}/\rho_{\rm halo}-1$ (the error was estimated inside $50\kpch$
radius); (11) the same as in the previous column, but for the fits of
profile advocated by Moore et al.

Halo $A$ in the first set of simulations was re-simulated three times with
increasing mass resolution. For each simulation, we considered outputs
at four time moments in the interval to $z=0-0.03$.
Parameters of the halos in these simulations averaged over
the four moments are presented in the first three rows of the Table~2.
We do not find any systematic change with resolution in the values of halo 
parameters both on the virial radius scale
and around the maximum of the circular velocity ($r=(30-40)\kpch$). 

\begin{figure}[tb!]
\epsscale{1.05}
\plottwo{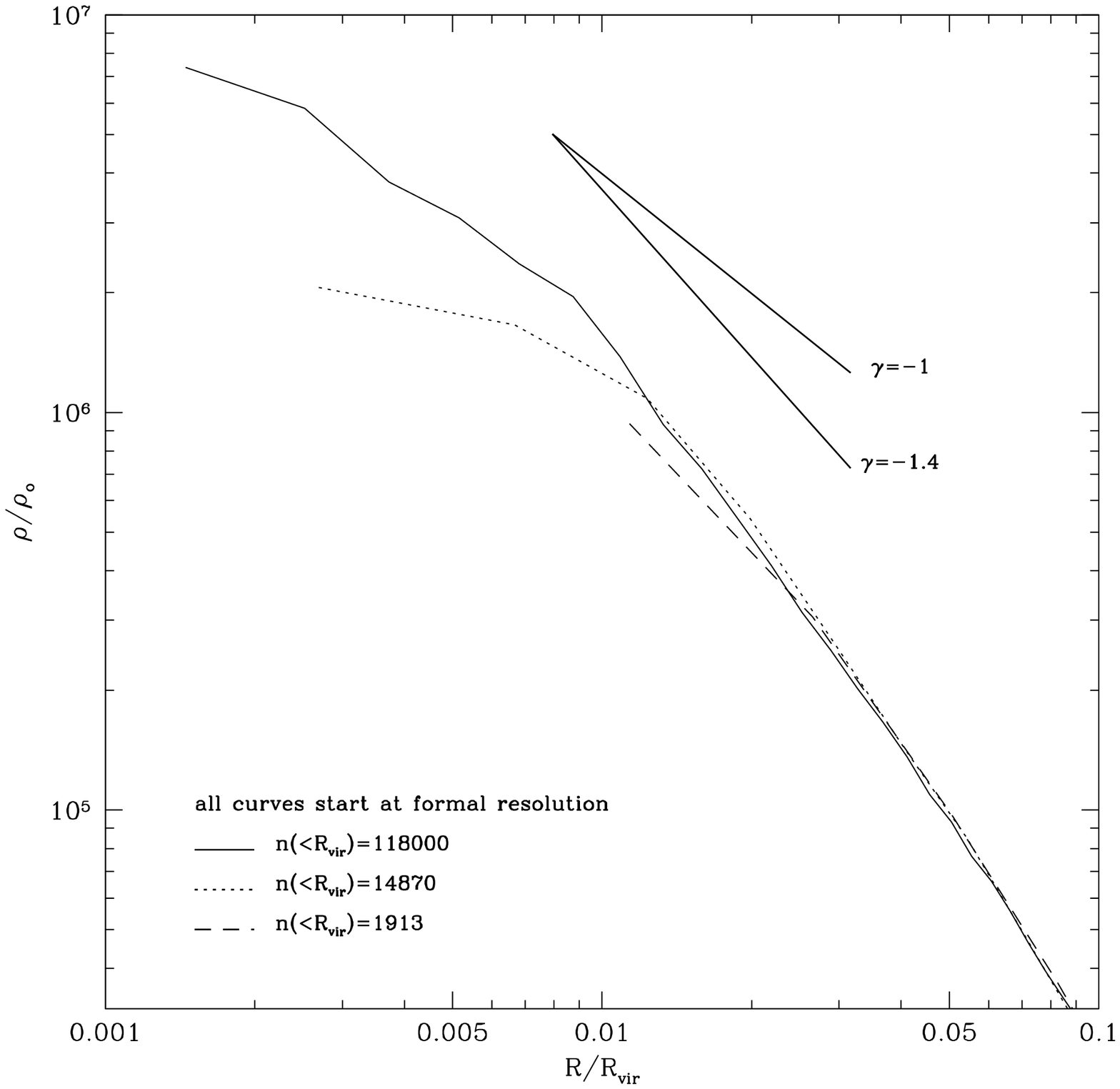}{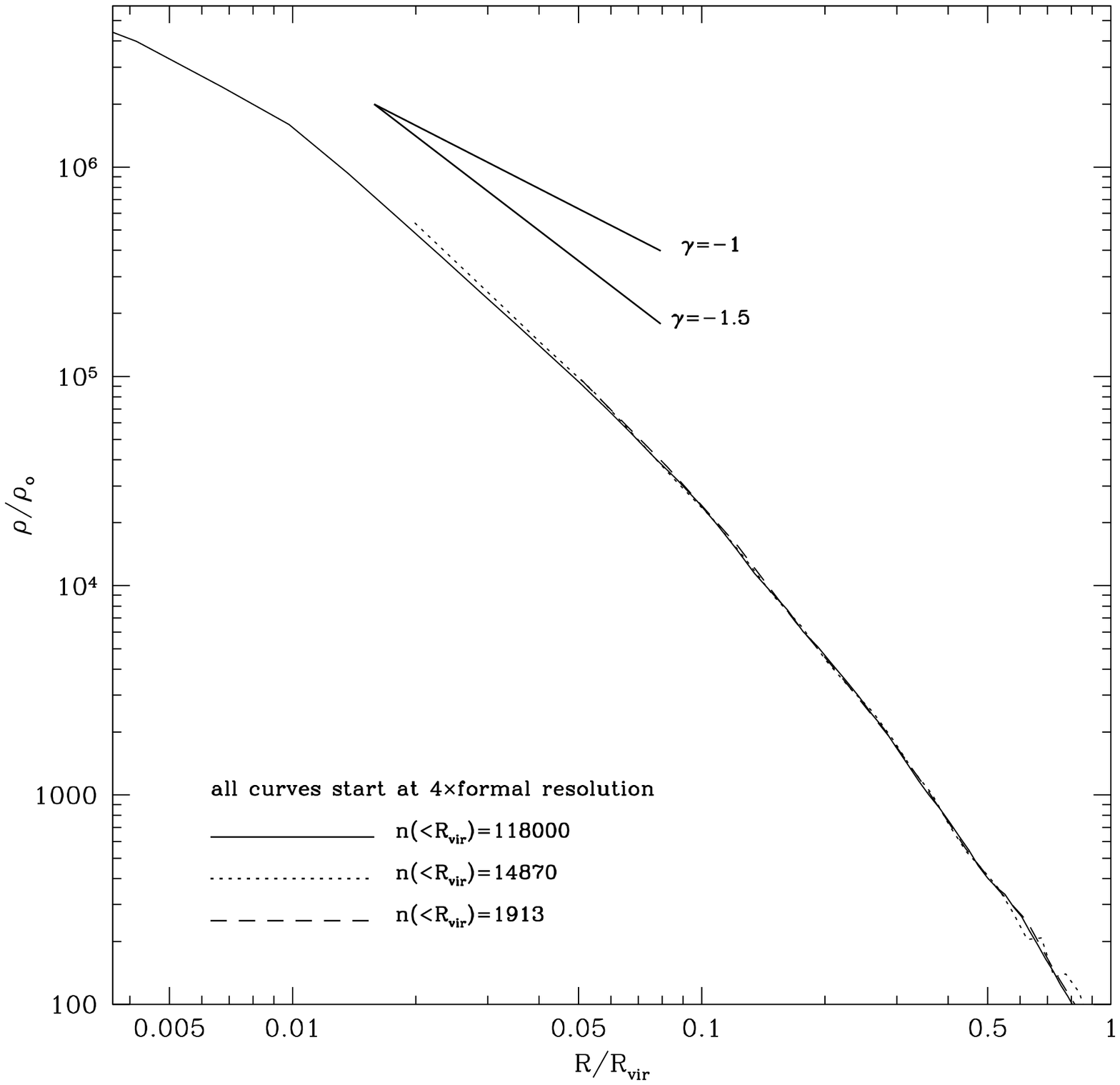}
\caption{\small {\it Left panel} Density profiles of halo  A simulated with different
mass and force resolutions. The profiles are plotted down 
to the formal force resolution of each simulation.{\it Right panel} The profiles
plotted down to {\em four formal resolutions}. It is clear that for vastly different 
mass (from 2000 to 120000 particles in the halo) and force (from $3.66\kpch$ 
to $0.23\kpch$) resolutions the convergence is reached at these scales. 
 }\label{fig:Benstyle}
\end{figure}

Left panel in Figure~\ref{fig:Benstyle} shows the central region of the halo A$_1$
(see Table~2). This plot is similar to the Fig.1a in Moore et al.
(1998) in that all profiles are drawn to the formal force resolution.
The straight lines indicate slopes of two power-laws: $\gamma=-1$
and $\gamma=-1.4$. The figure indeed shows that at around 1\% of the
virial radius the slope is steeper than -1 and the central slope
increases as we increase the mass resolution. Moore et al. (1998)
interpreted this behavior as evidence that profiles are steeper than
predicted by the NFW profile.  We also note that the results of our
highest resolution run $A_1$ are qualitatively consistent with results
of Kravtsov et al. (1998). Indeed, if the profiles are considered down
to the scale of {\em two} formal resolutions, the density profile slope in
the very central part of the profile $r\lesssim 0.01r_{\rm vir}$ is
close to $\gamma=-0.5$. 

The profiles in Figure~\ref{fig:Benstyle} reflect the density
distribution in the cores of simulated halos. However, the
interpretation of these profiles is not straightforward because it
requires assessment of numerical effects.  The formal resolution
usually does not even correspond to the scale where the numerical
force is fully Newtonian (usually it is still considerably ``softer''
than the Newtonian value).  In the ART code, the
interparticle force reaches (on average) the Newtonian value at about
two formal resolutions(see Kravtsov et al. 1997).  The effects of
force resolution can be studied by resimulating the same objects with
higher force resolution and comparing the density profiles. Such
convergence study was done in Kravtsov et al. (1998) where it was
found that {\em for a fixed mass resolution} halo density profiles
converge at scales above two formal resolutions.  Second, local
dynamical time for particles moving in the core of a halo is very
short. For example, particles on the circular orbit of the radius
$1\kpch$ from the center of halo $A$ makes about 200 revolutions over
the Hubble time. Therefore, if the time step is insufficiently small,
numerical errors in these regions will tend to grow especially fast.
The third possible source of numerical errors is mass resolution. Poor
mass resolution in simulations with good force resolution may, for
example, lead to two-body effects (e.g., Knebe et al. 2000).
Insufficient number of particles may also result in ``grainy''
potential in halo cores and thereby affect accuracy of orbit
integration. In these effects, the mass resolution may be closely
inter-related with force resolution.

It is clear thus that in order to make conclusions not affected by
numerical errors, one has to determine the range of trustworthy scales
using convergence analysis. Right panel in Figure~\ref{fig:Benstyle} shows that for
the halo A simulations the convergence for vastly different mass and
force resolution is reached for scales $\gtrsim 4$ formal force
resolutions (all profiles in this figure are plotted down to the
radius of 4 formal force resolutions).  For all resolutions, there are
more than 200 particles within the radius of four resolutions from the
halo center. For the highest resolution simulation (halo A$_1$) the
convergence is reached at scales $\gtrsim 0.005\rvir$.

\begin{figure}[tb!]
\epsscale{0.75}
\plotone{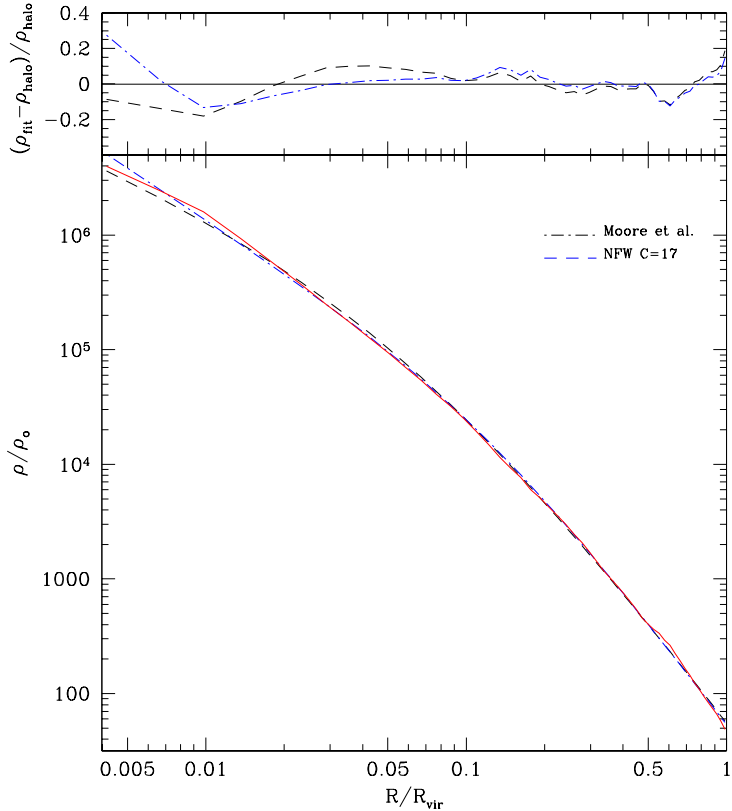}
\caption{\small  Fits of the NFW and Moore et al. halo profiles
to the profile of halo A$_1$ ){\em bottom panel}. The {\em top panel} shows
fractional deviations of the analytic fits from the numerical profile. 
Note that both analytic profiles fit numerical profile equally well: fractional
deviations are smaller than 20\% over almost three decades in radius. } \label{fig:FitsConverge}
\end{figure}

In order to judge which profile provides a better description of the
simulated profiles we fitted the NFW and Moore et al. analytic
profiles.  Figure~\ref{fig:FitsConverge} presents results of the fits
and shows that both profiles fit the numerical profile equally well:
fractional deviations of the fitted profiles from the numerical one
are smaller than 20\% over almost three decades in radius. It is clear
thus that the fact that numerical profile has slope steeper than $-1$
at the scale of $\sim 0.01\rvir$ does not mean that good fit of the
NFW profile (or even analytic profiles with shallower asymptotic
slopes) cannot be obtained.

\begin{figure}[tb!]
\epsscale{0.9}
\plotone{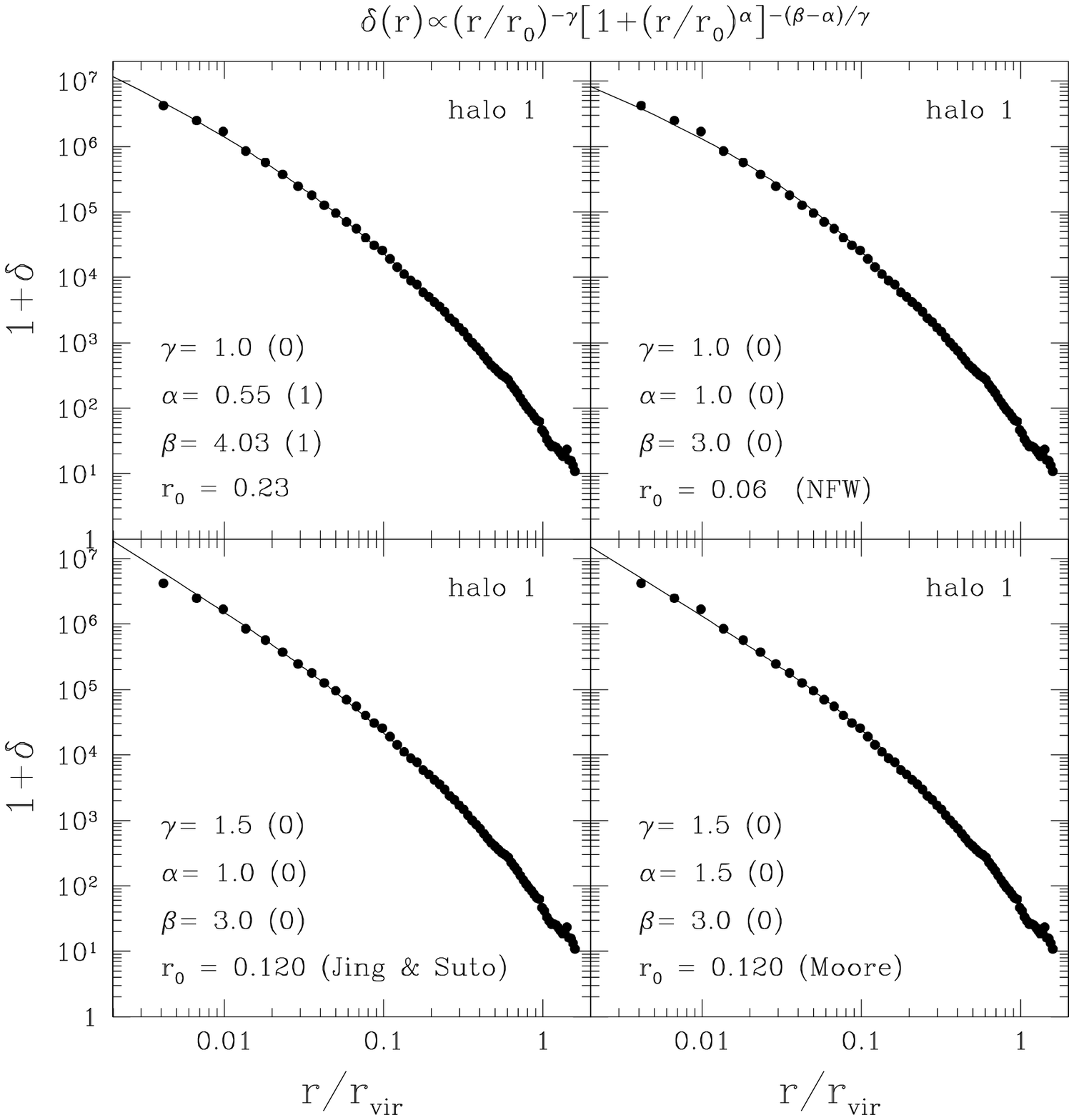}
\caption{\small Analytic fits to the density profile of the halo $A_1$ (see Table~2) from our set of simulations. The fits are 
  of the form $\rho(r)\propto (r/r_0)^{-\gamma}
  [1+(r/r_0)^{\alpha}]^{-(\beta-\alpha)/\gamma}$.  The legend in each
  panel indicates the corresponding values of $\alpha$, $\beta$, and
  $\gamma$ of the fit; the digit in parenthesis indicates whether the
  parameter was kept fixed ($0$) or not ($1$) during the fit. Note
  that various sets of parameters $\alpha$, $\beta$, $\gamma$ provide
  equally good fits to the simulated halo profile in the whole range
  resolved range of scales $\approx 0.005-1\rvir$. This indicates 
  a large degree of degeneracy in parameters $\alpha$, $\beta$, and
  $\gamma$} \label{fig:rf4_1}
\end{figure}

There is certainly a certain degree of degeneracy in fitting various
analytic profile to numerical results. Figure~\ref{fig:rf4_1}
illustrates this further by showing results of fitting profiles (solid
lines) of the form $\rho(r)\propto
(r/r_0)^{-\gamma}[1+(r/r_0)^{\alpha}]^{-(\beta-\alpha)/\gamma}$ to
{\em the same} (halo $A_1$) simulated halo profile shown as solid
circles. The legend in each panel indicates the corresponding values
of $\alpha$, $\beta$, and $\gamma$ of the fit; the digit in
parenthesis indicates whether the parameter was kept fixed ($0$) or
not ($1$) during the fit.  The right two panels show fits of the NFW
and Moore et al. profiles; the bottom left panel shows fit of the
profiles used by Jing \& Suto (2000). The top left panel shows a fit
in which the inner slope was fixed but $\alpha$ and $\beta$ were fit.
The figure shows that all four analytic profiles can provide a nice
fit to the numerical profile in the whole range $0.005-1\rvir$.

%==================================
\subsection{Halo profiles at $z=1$}
%==================================

As we mentioned in \S~\ref{sec:simulations}, the halo A analyzed in
the previous section is somewhat special because it was selected as an
isolated relaxed halo. In order to reach unbiased conclusions, in this
section we will present analysis of halos from the second set of
simulations (halos B, C, and D in Table~2) which were not selected to
be relaxed or isolated.  Based on the results of convergence study
presented in the previous section, we will consider profiles of these
halos only at scales above four formal resolutions use results
starting only from 4 formal resolutions and not less than 200
particles. Note that these conditions are probably more stringent than
necessary because these halos were simulated with $5-7$ times more
particles per halo. There is an advantage in analyzing halos at a
relatively high redshift. Halos of a given mass will have lower
concentration (see Bullock et al. 2000). Lower concentration implies a
large scale at which the asymptotic inner slope is reached. Profiles
of the high redshift halos should therefore be more useful in
discriminating between the analytic models with different inner
slopes.

\begin{figure}[tb!]
\epsscale{0.75}
\plotone{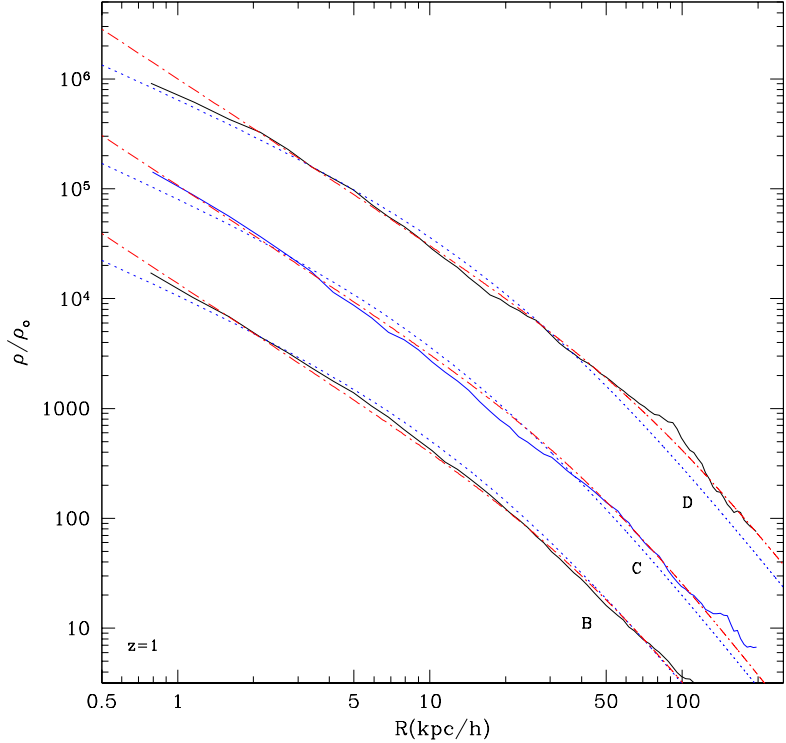}
\caption{\small Profiles of halos B, C, D at $z=1$. 
  Profiles of halos C and D were offset downwards by factors of 10 and
  100 for clarity. {\em Solid curves} show simulated profiles, while
  {\em dotted and dot-dashed curves} show the NFW and Moore et al.
  fits, respectively. The halo profiles in simulations are plotted
  down to four formal resolution. Each halo had more than 200
  particles inside the smallest plotted scale.}\label{fig:Profz}
\end{figure}

We found that substantial substructure is present inside the virial
radius in all three halos. Figure~\ref{fig:Profz} shows profiles of
these halos at $z=1$. There profiles are not as smooth as that of halo
A$_1$ due to the substructure. Note that bumps and depressions visible
in the profiles cannot are significantly larger amplitude than the
shot noise.  Halo $C$ appeared to be the most relaxed of the three
halos.  This halo had the last major merger somewhat earlier than the
other two.  Halo $D$ had a major merger event at $z\approx 2$. Remnant
of the merger are still visible as a hump at radii around $100\kpch$.
The non-uniformities of profiles cause by substructure may
substantially bias analytic fits to the entire range of scales below
the virial radius.  Therefore, we used only the central, presumably
more relaxed, regions in the analytic fits: $r<50\kpch$ for halo D and
$r<100\kpch$ for halos B and C (fits using only central $50\kpch$ did
not change results).

The best fit parameters were obtained by minimizing the maximum
fractional deviation of the fit: ${\rm max}({\rm abs}(\log\rho_{\rm
  fit})-\log\rho_{\rm halo})$. Minimizing the sum of squares of
deviations ($\chi^2$), as is often done, can result in larger errors
at small radii with the false impression that the fit fails because it
has a wrong central slope. The fit that minimizes maximum deviations,
improves the NFW fit for points in the range of radii $(5-20)\kpch$,
where the NFW fit would appear to be below the data points if the fit
was done by the $\chi^2$ minimization.  This improvement comes at the
expense of few points around $1\kpch$.  For example, if we fit halo B
by using $\chi^2$ minimization, the concentration decreases from 12.3
(see Table~2) to 11.8 We also made a fit for halo B assuming even more
stringent limits on the effects of numerical resolution. By minimizing
the maximum deviation we fitted the halo starting at six times the
formal resolution. Inside this radius there were about 900 particles.
Resulting parameters of the fit were close to those in Table~2:
$C_{\rm NFW}=11.8$, and maximum error of the NFW fit was 17\%.

We found that the errors in the Moore et al. fits were systematically
smaller than those of the NFW fits, though the differences were not
dramatic. Moore et al. fit failed for halo $D$. It formally gave very
small errors, but this was done for a fit with unreasonably small
concentration $C=2$. When we constrained the approximation to have
about twice larger concentration as compared with the best NFW fit, we
were able to obtain a reasonable fit (this fit is shown in
Figure~\ref{fig:Profz}). Nevertheless, the central part was fit poorly
in this case.

Our analysis therefore failed to determine which analytic profile
provides a better description of the density distribution in simulated
halos. Despite the larger number of particles per halo and lower
concentrations of halos, results are still inconclusive.  Moore at al.
profile is a better fit to the profile of halo C; the NFW profile is a
better fit to the central part of halo D. Halo B represents an
intermediate case where both profiles provide equally good fits (similar
to the analysis of halo A). 

Note that there seem to be  real deviations in parameters
of halos of the same mass. Halos B and D have the same virial radii
and nearly the same circular velocities, yet their concentrations are
different by 30\%. We find the same differences in estimates of $C_{1/5}$
concentrations, which do not depend on specifics of an analytic fit.
The central slope at around $1\kpc$ also changes from halo to halo. 

\section{Summary}

We run a series of simulations with vastly different mass and force
resolution with the goal of studying the shape of the dark matter halo
profile in central parts of galaxy-size halos. We use a modified
version of the ART code, which is capable of handling particles with
different masses, variable force, and time resolution. In runs with the
highest resolution, the code achieved (formal) dynamical range of
$2^{17}=131,072$ with 500,000 steps for particles at highest level of
resolution.

Our conclusions regarding the convergence of the profiles differ
from those of Moore et al. (1997). If we take into account only radii, at which we
believe numerical effects (the force resolution, the resolution of initial
perturbations, and the two-body scattering) are small, then we find that
the slope and the amplitude of the density do not change when we change
the force and mass resolution. This result is consistent with what was
found in simulations of the ``Santa Barbara'' cluster (Frenk et al.,
1999): at a fixed {\it resolved} scale results do not change as the
resolution increases. For the ART code the results
converged at 4 times the formal force resolution and more than 200
particles.  These limits of convergence very likely depend on
particular code used and on the duration of integration.

We reproduce results of Moore et al. regarding the convergence and
results of Kravtsov et al. (1998) regarding shallow central profiles,
but only when we consider points inside unresolved scales. We conclude
that those results followed from overly optimistic interpretation of
numerical accuracy of simulations.

For the galaxy-size halos considered in this paper  with masses $M_{\rm vir}
=7\times 10^{11}\Msunh - 2\times 10^{12}\Msunh$ and concentrations
$C=9-17$ both the NFW profile $\rho\propto r^{-1}(1+r)^{-2}$ and the
Moore et al. profile $\rho\propto r^{-1.5}(1+r^{1.5})^{-1}$ give good
fits with accuracy about 10\% for radii not smaller than 1\% of the
virial radius. None of the profiles is significantly better then the
other.  

Halos with the same mass may have different profiles. No matter what
profile is used -- NFW or Moore et al. -- there is no universal
profile: just halo mass does not yet define the density
profile. Nevertheless, the universal  profile is extremely useful
notion which should be interpreted  as the general trend $C(M)$ of halos with
larger mass to have lower concentration. Deviations from the general
$C(M)$ are real and significant \citep{Bullocketal99}.
It is not yet clear, but seems very likely that the central slopes of
halos also have real fluctuations. The fluctuations in the
concentration and the central slopes are important for interpretation
of the central parts of rotation curves.
 
\acknowledgements

We acknowledge the support of the grants NAG- 5- 3842 and NST-
9802787. A.V.K. acknowledges support by NASA through Hubble Fellowship
grant HF-01121.01-99A from the Space Telescope Science Institute,
which is operated by the Association of Universities for Research in
Astronomy, Inc., under NASA contract NAS5-26555.  Computer simulations
presented in this paper were done at the National Center for
Supercomputing Applications (NCSA), Urbana-Champaign, Illinois.

%===================================================================

\end{document}